\pdfoutput=1
 \documentclass[final,3p,times,twocolumn]{elsarticle}
\usepackage{amssymb}
\usepackage{bm}
\usepackage{amsmath}
\usepackage{xcolor}
\usepackage{lineno}
\usepackage{comment}
\usepackage{graphicx}
\usepackage[caption=false]{subfig}
\journal{Nuclear Instruments and Methods in Physics Research A}

\begin{document}

\begin{frontmatter}

\title{Measurement of the response function of the PIENU calorimeter}

\author[a]{A. Aguilar-Arevalo}
\author[b]{M. Aoki}
\author[c]{M. Blecher}
\author[d]{D.I. Britton}
\author[e,f]{D.A. Bryman}
\author[f]{L. Doria\fnref{fn1}}
\author[g]{S. Cuen-Rochin}
\author[f]{P. Gumplinger}
\author[a]{I. Hernandez}
\author[f,h]{A. Hussein}
\author[b]{S. Ito\fnref{fn2}}
\author[f]{L. Kurchaninov}
\author[i]{L. Littenberg}
\author[e,f,j]{C. Malbrunot}
\author[f]{R.E. Mischke}
\author[f]{T. Numao}
\author[d]{D. Protopopescu}
\author[f]{A. Sher}
\author[e]{T. Sullivan\fnref{fn3}}
\author[f]{D. Vavilov}

\address[a]{Instituto de Ciencias Nucleares,
            Universidad Nacional Autonoma de Mexico,
            04510,
            CDMX,
            Mexico}
            
\address[b]{Graduate School of Science,
            Osaka University,
            Toyonaka,
            560-0043,
            Osaka,
            Japan}

\address[c]{Physics Department,
            Virginia Tech.,
            Blacksburg,
            24061,
            VA,
            USA}

\address[d]{Physics Department,
            University of Glasgow,
            Glasgow,
            U.K.}

\address[e]{Department of Physics and Astronomy,
            University of British Columbia,
            Vancouver,
            V6T 1Z1,
            B.C.,
            Canada}

\address[f]{TRIUMF,
            4004 Wesbrook Mall,
            Vancouver,
            V6T 2A3,
            B.C.,
            Canada}

\address[g]{Tecnologico de Monterrey,
            Escuela de Ingenieria y Ciencias,
            Culiacan,
            80100,
            Sinaloa,
            Mexico}

\address[h]{University of Northern British Columbia,
            Prince George,
            V2N 4Z9,
            B.C.,
            Canada}

\address[i]{Brookhaven National Laboratory,
            Upton,
            11973-5000,
            NY,
            USA}

\address[j]{McGill University,
            Montr\'{e}al, 
            H3A 2T8, 
            Qu\'{e}bec,  
            Canada}

\fntext[fn1]{Present address: Institut für Kernphysik,
            Johannes Gutenberg-Universität Mainz,
            Mainz,
            55122,
            Germany}

\fntext[fn2]{Present address: National Institute of Technology,
            Kitakyushu College,
            Kitakyushu,
            802-0985,
            Kokuraminami,
            Fukuoka,
            Japan}

\fntext[fn3]{Present address: University of Victoria,
            Victoria,
            V8P 5C2,
            B.C.,
            Canada}

\begin{abstract}

\noindent Measurements of the response function of the PIENU NaI(T$\ell$) and CsI crystal calorimeter using  a monochromatic 70 MeV/c positron beam at various incidence angles are described. The experimental setup and relevant physical processes involved were simulated using Geant4 to reproduce positron energy spectra. Monte Carlo simulations were compared with experimental data across ten calorimeter-beam angles and showed good agreement.
%within a 1$\sigma$ uncertainty band. 
This allowed the validation of  simulation parameters that were essential for precise  measurements of pion decays.

\end{abstract}

\begin{keyword}
positron beam \sep electromagnetic shower \sep pion decay \sep response function
\end{keyword}

\end{frontmatter}

\section{Introduction}

The PIENU experiment \cite{PIENU_exp} aims to measure the pion decay branching ratio, $R_{e/\mu} = \Gamma(\pi\rightarrow e\nu+\pi\rightarrow e\nu\gamma)/\Gamma(\pi\rightarrow \mu\nu + \pi\rightarrow \mu\nu\gamma$) to a precision of $\mathcal{O}$(0.1$\%$). This ratio is critical for understanding pion decay dynamics and serves as a key observable for testing predictions of the Standard Model and constraining potential new physics \cite{PBSM}. The ratio $g_{\mu}/g_{e}$ of muon and electron weak interaction coupling constants can be obtained from measurement of $R_{e/\mu}$ to test the principle of lepton flavor universality.

The PIENU technique involved stopping a pion beam in an active scintillation target from which decay positrons emerged at various angles and interacted with a calorimeter. The calorimeter \cite{PIENU_cal} featured a single-crystal NaI(T$\ell$) detector for the positron energy measurement, surrounded by an array of CsI crystals to mitigate the energy loss by electromagnetic shower leakage. The response function of the calorimeter depended on the angle of incidence of the positrons which  determined the amount of material along the particle path and the amount of leakage from the electromagnetic shower.

The PIENU detector was designed to separate the rare $\pi^+\rightarrow e^+ \nu_e$  decay and the $\pi\rightarrow \mu\rightarrow e$ decay chain: $\pi^+ \rightarrow \mu^+ \nu_\mu$ followed by $\mu^+ \rightarrow e^+ \overline{\nu{_\mu}} \nu_e$ decays. In the former, a charged pion decays to a monoenergetic positron of total energy $E_e=69.8$ MeV and an electron neutrino with the branching ratio $R_{e/\mu} \approx 10^{-4}$.  The $\pi\rightarrow \mu\rightarrow e$ decay chain  produces a positron energy spectrum up to $E_e=52.8$ MeV. To separate the two decay channels the calorimeter energy spectrum is divided into two parts: the high energy part includes only $\pi\rightarrow e\nu$ decays, while the low energy part includes all $\pi\rightarrow \mu\rightarrow e$ events plus a fraction of $\pi\rightarrow e\nu$ events, which are buried under the muon decay spectrum.

The tail fraction is the number of $\pi\rightarrow e\nu$ events below an energy cut divided by the total number of events of the $\pi\rightarrow e\nu$ energy spectrum. In the PIENU experiment, the low energy tail is produced mainly by Bhabha scattering, energy leakage from the electromagnetic shower, and pion radiative decays \cite{PIENU_exp}. The tail fraction is the largest correction to $R_{e/\mu}$ and is determined via a Monte Carlo (MC) simulation since it cannot be measured directly. 

Given the interactions in the low energy regime, the tail fraction serves as a mean to quantify the response function of the PIENU calorimeter. To measure the tail fraction independently, positron beam energy spectra were taken at ten different incident angles. The amount of material in the positron path, an important feature for the response function, was determined by the angle of incidence to the calorimeter. The tail fraction of each positron beam spectrum was used to characterize the response function. The energy spectra were then compared to matching MC simulations at each angle. By confirming that the MC simulation could reproduce the positron beam data, confidence was gained for using MC to calculate the tail correction in the $R_{e/\mu}$ measurement and to reduce its systematic uncertainty. 

\section{Experimental setup}

The TRIUMF cyclotron produced a 520 MeV proton beam transported by a beam-line to a 1 cm thick Beryllium production target. Particles emerging from the target entered the secondary channel M13 \cite{PION_BEAM}, a low-momentum achromatic beam line. M13 was tuned to obtain a positron beam with momentum 70$\pm$1 MeV/c which was delivered to the PIENU detector. The beam composition was approximately 63\% $\pi^+$, 11\% ~$ \mu^+$, and 26\%$ ~e^+$. The time of arrival of particles at the detector with respect to the 23 MHz cyclotron radio-frequency time (TOF) and the energy losses in the scintillation counters provided particle identification \cite{OramM13}.

\begin{figure}[t!]
  \begin{center}
    \includegraphics[width=0.48\textwidth]{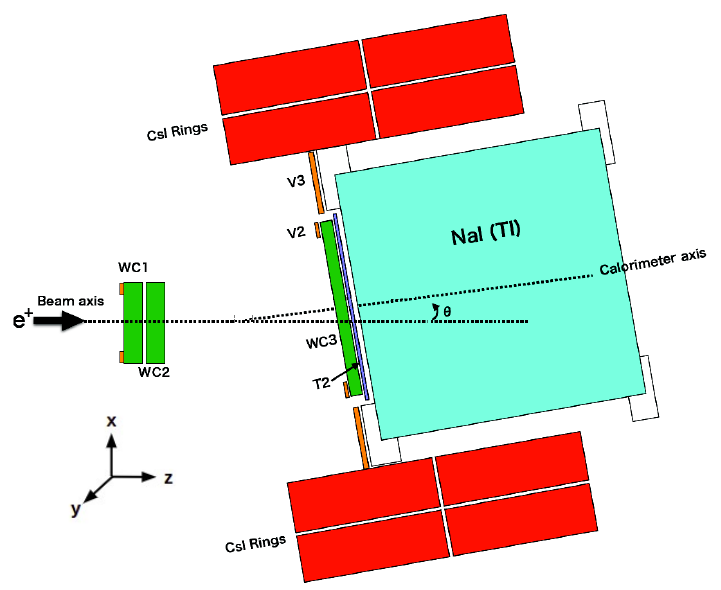}
    \caption{Schematic drawing of the Lineshape setup for special positron runs, showing the angle of rotation ($\theta$) between the beam and calorimeter. See the text for details.}
  \label{exp_setup}
  \end{center}
\end{figure}

To study the effect of  positrons hitting the front face of the calorimeter at various angles, several components of the PIENU detector used for the $R_{e/\mu}$ measurement were removed including some detectors used for pion identification and tracking. This configuration  facilitated rotation of the calorimeter with respect to the beam axis. The experimental configuration shown in Figure \ref{exp_setup} will be referred to as the Lineshape setup. The wire chambers WC1 and WC2 provided incident beam position measurements. Each wire chamber, along with WC3 (located near the front face of the NaI(T$\ell$) crystal), consisted of three wire planes that were rotated by an angle of 120$^\circ$ with respect to each other. Each wire plane of WC1 and WC2 consisted of 120 wires (0.8 mm pitch), grouped in threes and connected to a read-out pad. Each plane of WC3 had 96 wires, grouped in twos for an effective pitch of 4.8 mm. The active diameters were 9.6 cm for WC1 and WC2 and 23 cm for WC3. A 6.6-mm thick plastic scintillation detector (T2), located downstream of WC3, triggered the positron signal.

The calorimeter consisted of a 480-mm diameter and 480-mm long (19 radiation length) NaI(T$\ell$) detector housed in a 3-mm thick aluminum enclosure \cite{NaI_description} and viewed by 19 photomultipliers (PMT). The NaI(Tl) crystal was surrounded by 97 CsI crystals \cite{CsI_description} arranged in two upstream and two downstream concentric layers. Each concentric layer was supported by a 2 mm thick stainless steel cylinder. The resulting array of CsI crystals was 50 cm long and 16 cm thick (9 radiation lengths) in the radial direction. The PIENU calorimeter was equipped with four COPPER boards to digitize the signals from the analog sum of NaI(T$\ell$) PMTs, and partial analog sums of CsI array PMTs \cite{PIENU_det}. The crystal-beam angle ($\theta$) was measured using a theodolite by placing targets on the beampipe and on the cart on which the detector sat. Its precision was determined by measuring the relative position between the theodolite and the targets with high accuracy resulting in an angle uncertainty better than 0.1$^\circ$. The center of rotation was aligned with the center of the pion stopping target to replicate the pion decay angle. The alignment of the center of rotation was performed with a $\pm$ 1 mm precision.

\begin{table}[h]
\centering
\resizebox{0.47\textwidth}{!}{\begin{tabular}{cccccc}
\hline\hline
& & Data & & MC \\
Angle ($^\circ$) & Events (10$^{6}$) & Tail ($\%$) & Sys ($\%$) & Tail ($\%$) & Sys ($\%$) \\
\hline
0.0  &  5.97  & 0.54    & 0.02   & 0.55 & 0.11    \\
6.0  &  11.65  & 0.55    & 0.02   & 0.56 & 0.11    \\
11.8 &  6.51  & 0.58    & 0.02   & 0.58 & 0.11    \\
16.5 &  7.25  & 0.63    & 0.02   & 0.64 & 0.11    \\
20.9 &  6.53  & 0.71    & 0.02   & 0.72 & 0.11    \\
24.4 &  6.28  & 0.80    & 0.03   & 0.81 & 0.12    \\
30.8 &  5.78  & 1.14    & 0.04   & 1.11 & 0.12    \\
36.2 &  5.86  & 1.45    & 0.04   & 1.45 & 0.13    \\
41.6 &  6.43  & 2.12    & 0.06   & 2.09 & 0.14    \\
47.7 &  9.06  & 3.61    & 0.10   & 3.60 & 0.16    \\ \hline\hline      
\end{tabular}}
\caption{The number of events for each angle $\theta$ followed by the tail fractions for data and MC with systematic uncertainties. The statistical uncertainties were negligible.}
\label{systematic_muon}
\end{table}

To significantly reduce the background from pions in the beam, approximately half of the 43 ns RF period was vetoed in the trigger configuration. Additionally, a hit in the scintillator T2 was required in the trigger. Table \ref{systematic_muon} shows the number of events collected for each of the data-taking angles.

\vspace{0.1cm}

\subsection{Possible Low-momentum Beam Positrons}

Since the positrons had the potential to scatter in the beamline, the tail distribution could be contaminated by intrinsic low-momentum positrons coming from the M13 beamline. Although this possible low-energy component could not be directly identified,   the transport of positrons through the M13 channel using G4beamline \cite{G4Beamline} gave an upper limit to the low-energy tail contribution of (2.8 $\pm$ 0.5)$\times 10^{-4}$.  An additional constraint from a comparison of the data energy spectrum to MC will be mentioned later.

\section{Data analysis}

\subsection{Gain Stabilization and Energy calibration}

Each CsI PMT was connected to a Xe lamp through a quartz fiber. The pulse height of the Xe lamp signal in each PMT was compared run by run to a reference run in order to adjust for the PMT’s gain fluctuation.  The CsI crystal array was energy calibrated using cosmic rays via a dedicated cosmic trigger. The cosmic peak in each crystal was associated with the energy deposit predicted by a simulated cosmic shower. The reference run was renewed every 20 runs at the same time as the new cosmic calibration was obtained. 
In the case of the PMTs for the NaI(T$\ell$) crystal, automatic run-by-run gain correction was performed during data analysis by aligning the beam positron peak in each PMT with a reference run prior to being included in the calibrated sum.  The energy calibration of the NaI(T$\ell$) crystal is described in section 4.

\subsection{Event selection}

\begin{figure}[t!]
  \begin{center}
    \includegraphics[width=0.48\textwidth]{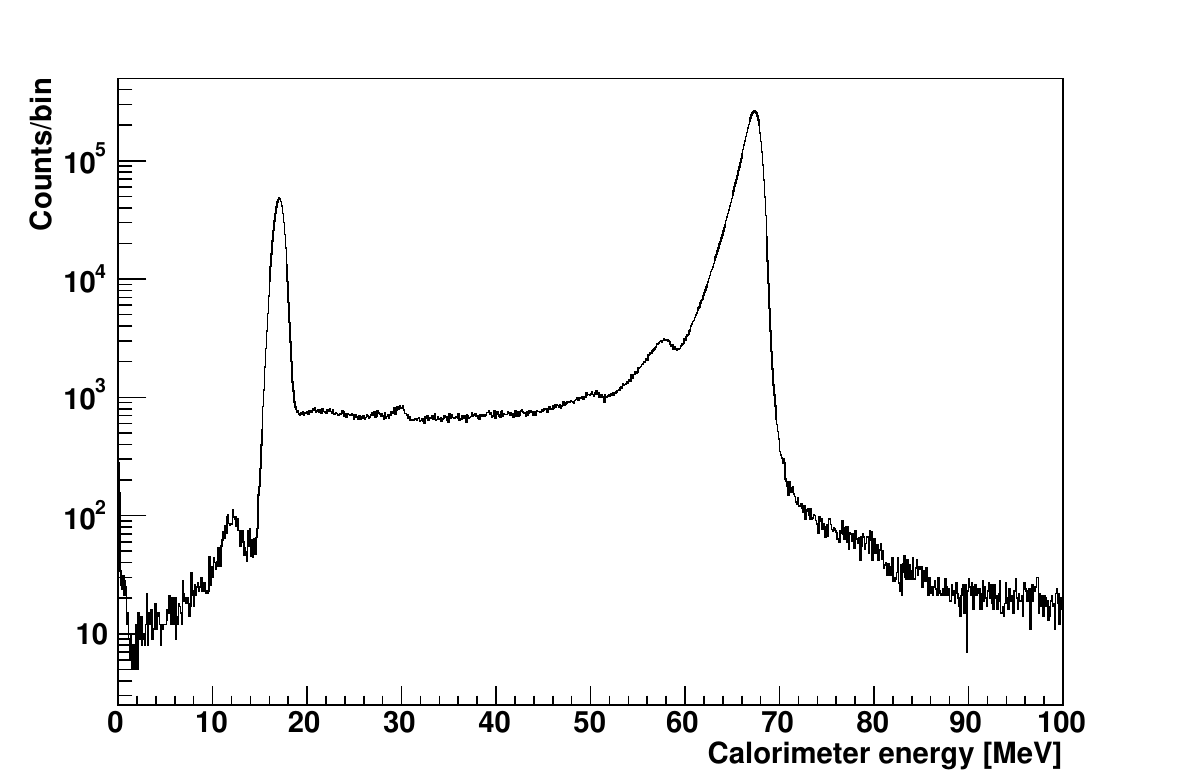}
    \caption{The energy spectrum in the PIENU calorimeter for $\theta=0^\circ$ after time and acceptance cuts in WC1 and WC2. The positron peak is located at 68 MeV and the remaining features are a small peak from beam pions around 13 MeV, a beam muon peak at 17 MeV, a slight bump at about 30 MeV caused by pion decays in flight, photonuclear peaks at 50 and 58 MeV \cite{PIENU_cal}, and a high energy tail above 70 MeV due to pileup from beam muons decaying in the NaI(T$\ell$) crystal.}
  \label{ES_cuts2}
  \end{center}
\end{figure}

Event selection was made using two stages of cuts to further remove background events. The first stage removed pileup and background events. For this, at least one hit in at least one plane of each wire chamber was required. Also, a timing cut in WC1 and WC2 was applied to reject events with hits outside a 100 ns window around the trigger time. To remove particles (mostly muons) from the beam halo, spatial acceptance cuts on the wire chambers WC1 and WC2 were required. The energy spectrum after the acceptance and timing cuts is depicted in Figure \ref{ES_cuts2}.

\subsection{Muon subtraction}

\begin{figure}[t!]
  \begin{center}
    \includegraphics[width=0.48\textwidth]{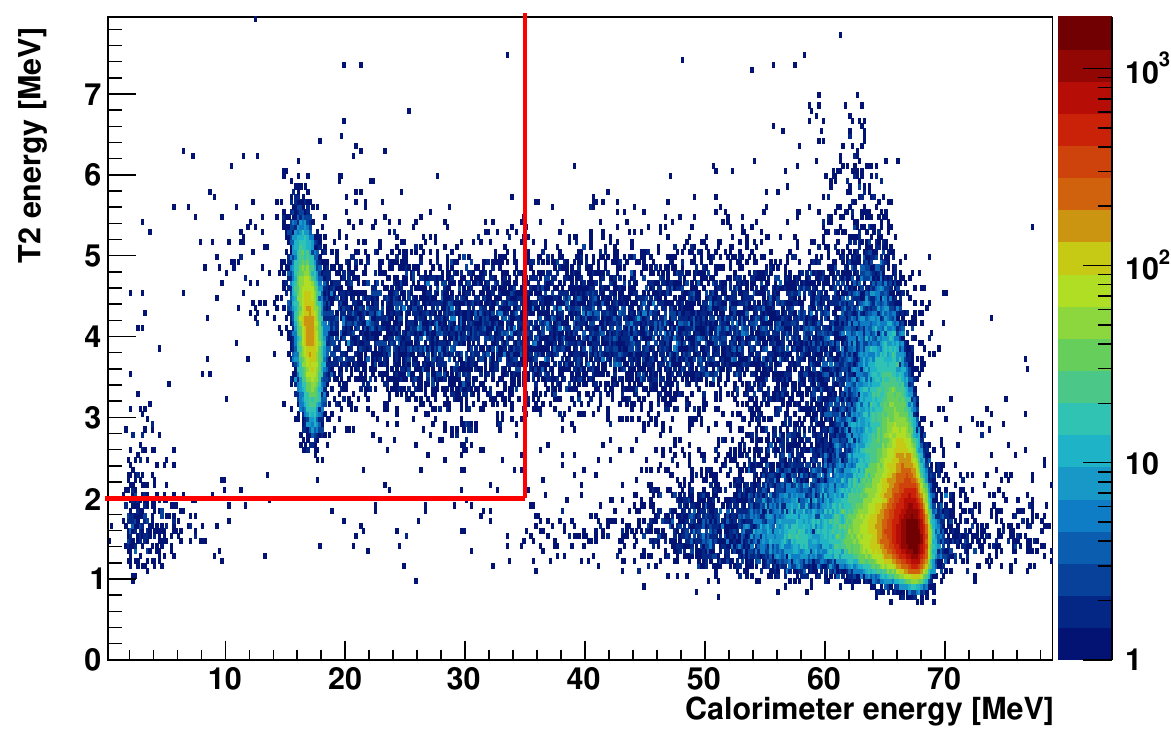}
    \caption{Energy deposit in T2 versus that in the calorimeter. The bright spot at 68 MeV calorimeter energy corresponds to positrons. It begins at about 1.1 MeV in T2 energy and extends to higher T2 energies due to the Landau distribution of energy loss in T2.  The spot at 17 MeV in calorimeter energy corresponds to muons. It is centered at about 4 MeV in T2 energy. Muons that decay in the calorimeter account for the stripe that extends from the muon spot up to $\sim$70 MeV in calorimeter energy. The red solid lines represent the 2D cut applied to remove most beam muons.}
  \label{NaI_T2}
  \end{center}
\end{figure}

The second stage consisted of two 2D cuts to remove beam muons. This was important for precise measurement of the tail fraction, and is illustrated for $\theta=0^\circ$ in Figure \ref{NaI_T2} which shows the 2D plot of energy in T2 versus calorimeter energy. The band centered at 1.6 MeV in T2 energy corresponded to positrons, while the band spreading from 2.5 to 6.5 MeV in T2 energy corresponded to beam muons. While a cut in T2 energy can reject most beam muons, it also removes positrons that have larger energy deposits in T2 due to the Landau distribution, and positrons from muon decay going backwards out of the NaI(T$\ell$) crystal. To mitigate the effect of a T2 energy cut on the positron spectrum, a two-dimensional cut in T2 energy versus calorimeter energy was employed. The cut removed events with energies in T2 higher than 2 MeV and calorimeter energy less than 35 MeV. These boundaries were set to remove most beam muons without compromising the tail fraction. The 2D cut caused a dip at 35 MeV in the low energy tail spectrum but the change in the fraction of positrons ($<0.001\%$) was negligible for determining the tail fraction.

\begin{figure}[t!]
  \begin{center}
    \includegraphics[width=0.48\textwidth]{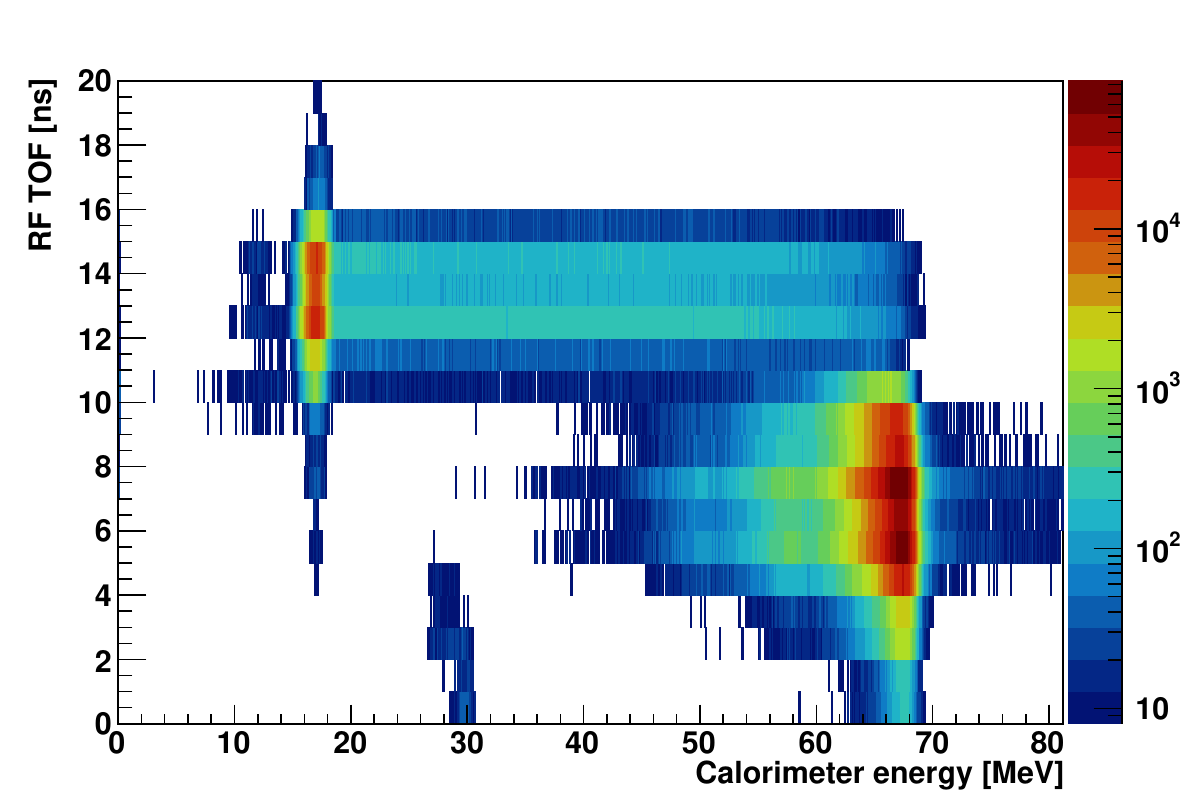}
    \caption{The time of flight (TOF) versus the calorimeter energy. The upper band (12 - 15 ns) corresponds to muons and the lower band (4 - 11 ns) corresponds to positrons.}
  \label{Ecal_TOF}
  \end{center}
\end{figure}

\begin{figure}[h!]
\centering 

\subfloat{%
  \includegraphics[clip,width=0.98\columnwidth]{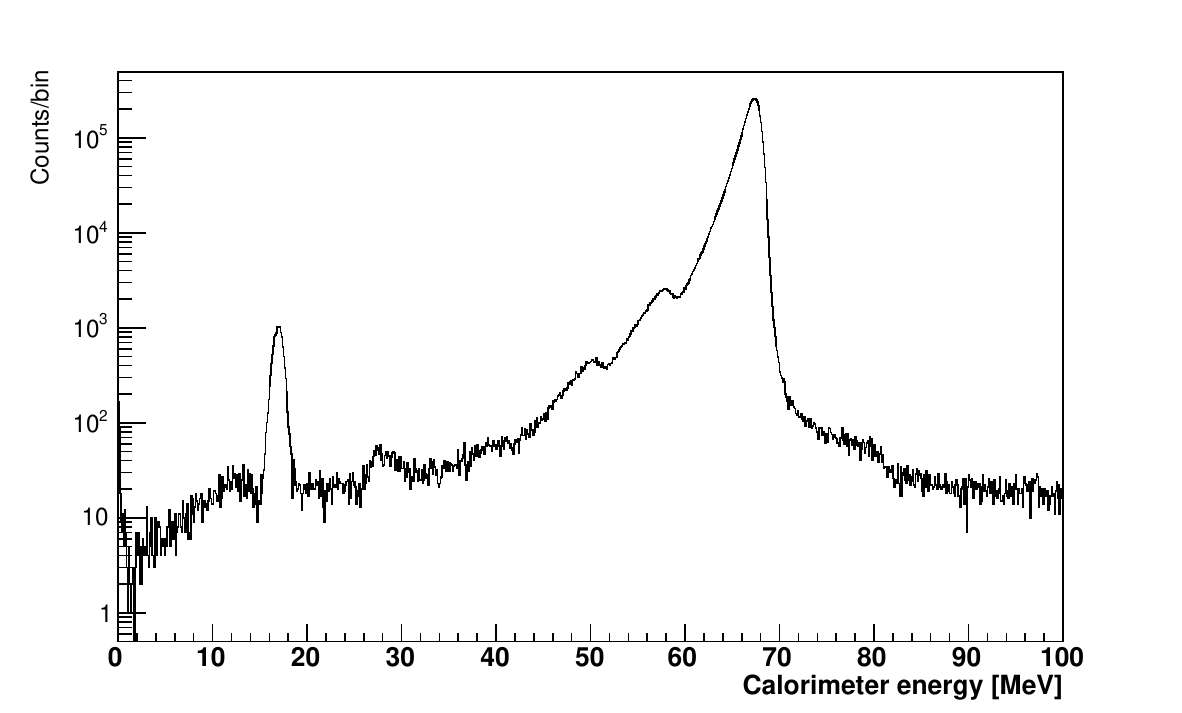}%
}

\subfloat{%
  \includegraphics[clip,width=0.98\columnwidth]{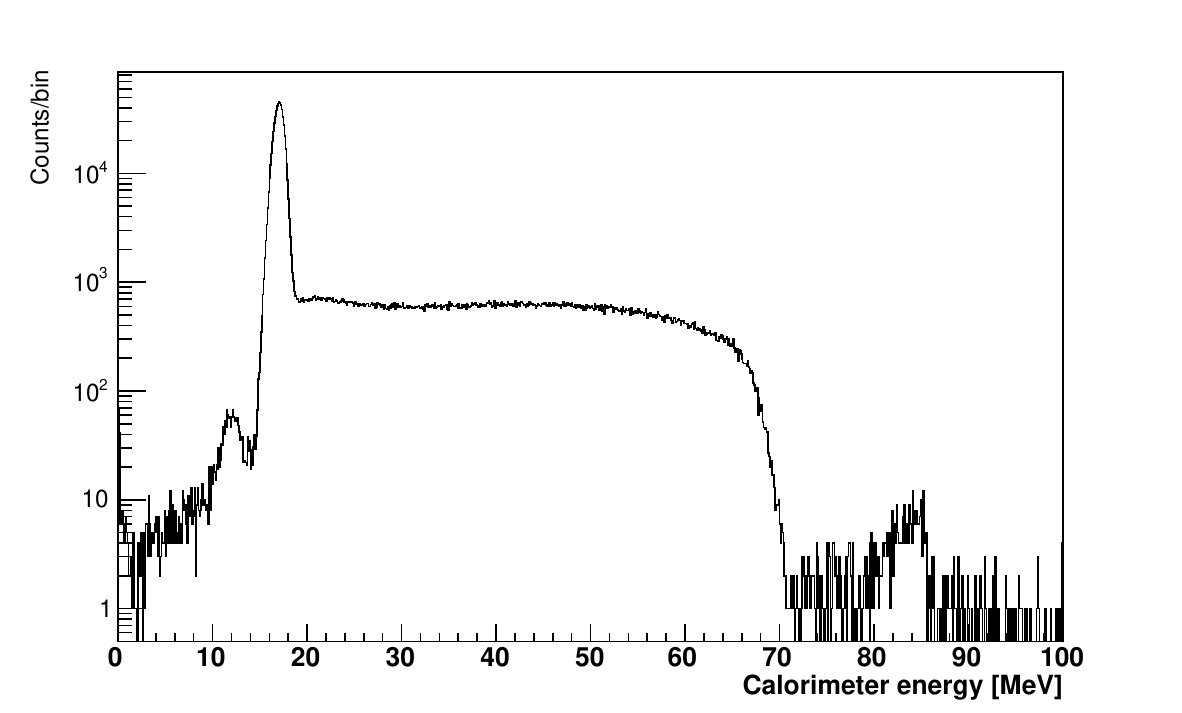}%
}

\subfloat{%
  \includegraphics[clip,width=0.98\columnwidth]{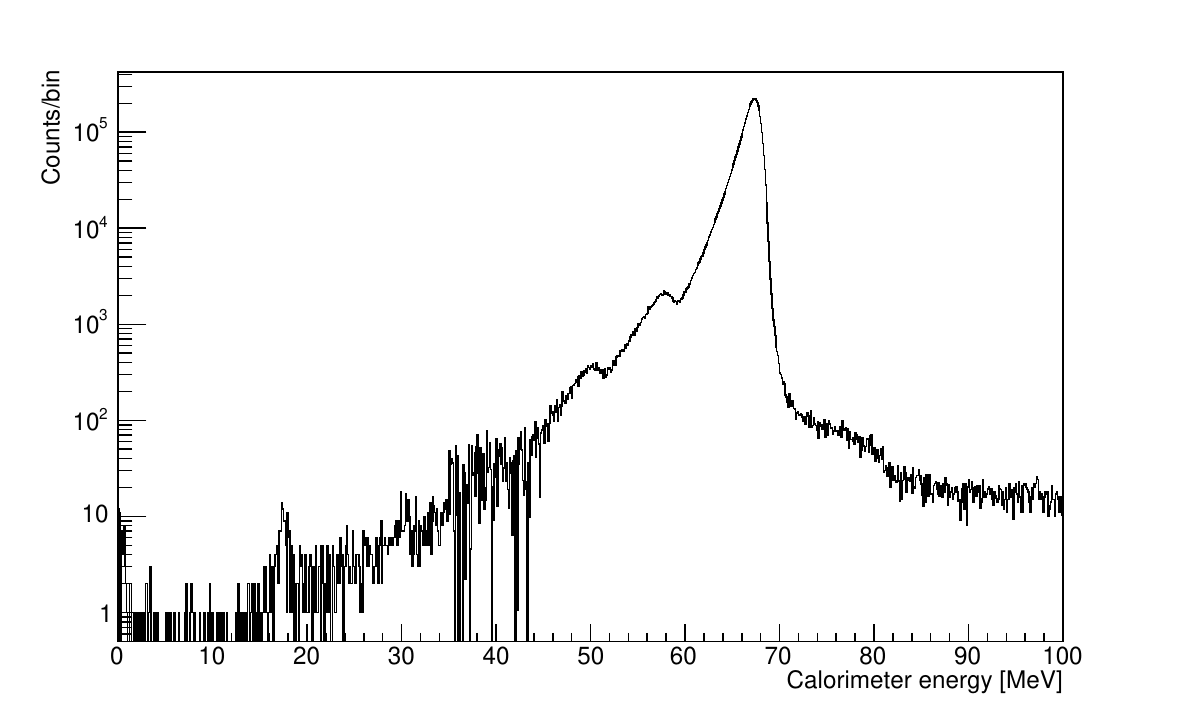}%
}
\caption{Upper: Positron energy spectrum in the calorimeter for $\theta=0^\circ$. Middle: Muon energy spectrum in the calorimeter. Both data portions were selected using RF TOF cuts. Lower: Positron energy spectrum after the 
 muon subtraction in the calorimeter. The drop at 35 MeV is due to the two-dimensional cut (see text).}
\label{ES_muon_pos}
\end{figure}

The remaining muon contribution above 35 MeV was subtracted using the known energy spectrum of muon events, selected by TOF. The correlation between TOF and the deposited energy in the calorimeter is depicted in Figure \ref{Ecal_TOF}. The energy spots are consistent with the peaks in Figure \ref{ES_cuts2}. By selecting 4-11 ns and 12-15 ns in TOF, the positron and muon energy spectra, shown in Figure \ref{ES_muon_pos}, were obtained with virtually no positron events remaining in the muon spectrum (middle plot in Figure \ref{ES_muon_pos}). By normalizing the muon peak in the muon spectrum to the remaining muon peak at 17 MeV in the positron spectrum, the muon contribution was subtracted. The resulting spectrum is shown in the lower plot in Figure \ref{ES_muon_pos}. The peaks at 50 and 58 MeV are due to photonuclear interactions \cite{PIENU_cal} followed by the emission and escape of one and two neutrons, respectively which may be induced by the positron's electromagnetic field or the emission of bremsstrahlung photons. For photons with energies close to the giant dipole resonance (GDR), the cross section for interacting with iodine nuclei in the NaI(Tl) is large. Frequently, one or more neutrons could be produced, and the probability of neutron escape from the calorimeter was significant. The high energy tail above 70 MeV was produced by pileup events. There is a remaining small portion of muons left in the 17 MeV peak and a small number of pion decaying-in-flight events at 30 MeV, confirmed by MC simulation. Both had negligible contributions to the tail fraction ($<$0.01$\%$). Table \ref{systematic_muon} shows the results of the tail fraction after the muon subtraction for a cut energy of 53.7 MeV. The uncertainty in the muon subtraction process was evaluated by varying the number of muons in the positron energy spectrum using different TOF cuts. The systematic uncertainty resulting from the muon subtraction process contributes to the uncertainties for data shown in Table \ref{systematic_muon}.

\section{Monte Carlo tuning}

The Monte Carlo simulation of the Lineshape setup was done with Geant4 version 10.06 \cite{Geant4}. The simulation included the detailed detector geometry \cite{PIENU_det} and a positron beam generated from the beam profiles measured by the wire chambers WC1 and WC2. These distributions were taken by sampling beam variables from data to accurately reproduce the beam characteristics. Because of changes in the beam conditions in the span of the data taking period, two beam profiles were generated. One beam profile was common to the first seven angles measured and another profile was used for the last three angles.

\begin{figure}[t!]
  \begin{center}
    \includegraphics[width=0.475\textwidth]{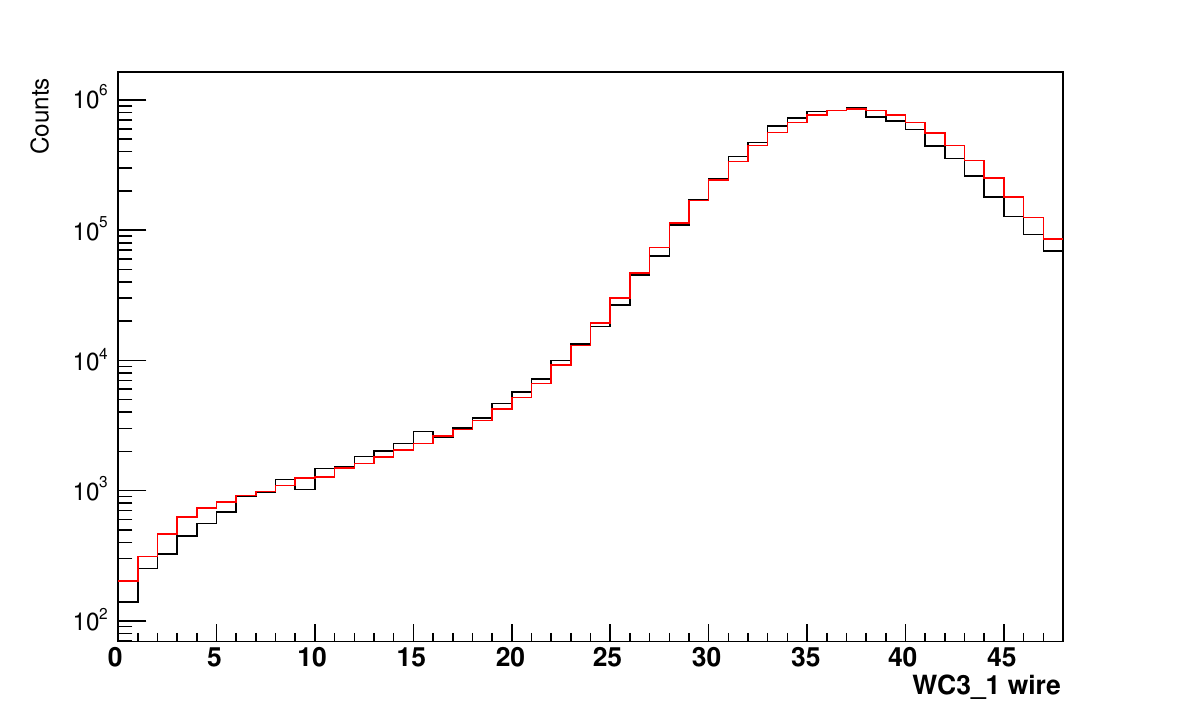}
    \caption{Distribution of activated wires in the first plane of WC3 for the set of data taken at $\theta=47.7^{\circ}$. Data are shown in black and MC in red.}
  \label{WC3_W_LS}
  \end{center}
\end{figure}

The beam distribution in the X and Y directions were smoothed and sampled randomly.  Due to the limited angular resolution of WC1 and WC2, the X and Y angular distributions were sampled from Gaussian distributions chosen to reproduce the spatial event distributions in WC3.  The correlation between these variables was encoded in a Cholesky matrix \cite{Cholesky} that was determined from correlations in the beam data. Tuning the angular distribution of the beam along the X axis and its divergence to match the data was especially important for the larger angles of rotation of the calorimeter as shower leakage became very important for these angles. The correction applied to the nominal beam angular distribution was typically below 1 mrad, with a largest correction of 12 mrad. Figure \ref{WC3_W_LS} depicts the wire distribution in the first plane of WC3 at 47.7$^{\circ}$ and shows good agreement between data and MC. The beam momentum distribution was modeled as a Gaussian with a central value of 70.0 $\pm$ 0.64 MeV/c.
%0.32 keV.

The energy spectrum of simulated beam positrons interacting with the calorimeter should ideally match the Lineshape data at each angle. To achieve this, choices were made among the Geant4 physics lists that best reproduced the data. In addition, three other adjustments were made to the MC to match the data.

For the choice of electromagnetic physics, it was found that agreement with data was achieved by choosing the PENELOPE (Penetration and Energy Loss of Positrons and Electrons) code system \cite{PENELOPE1}. PENELOPE was developed to contain considerably improved positron interaction models and more elaborate tracking algorithms for simulation of intermediate-energy positron transport involving complex geometries. The code generates electron–photon showers produced by primary particles (electrons, photons or positrons) in matter and operated from 100 eV up to 1 GeV. Its reliability has been verified with experimental data for nuclei with atomic number $Z = 1-92$ \cite{PENELOPE2}. In contrast, the default version of electromagnetic physics (EM0) for Geant4 10.06 yielded an energy spectrum that significantly disagreed with the Lineshape data.

The choice of hadronic physics list was made by comparing MC to data for the peaks at 58 and 50 MeV, which correspond to the escape of 1 or 2 neutrons, respectively, each with an energy of about 8 MeV. These features of the energy spectrum due to photonuclear interactions are important and are challenging to reproduce. Experimentally measured photonuclear cross-sections for the relevant target nuclei \cite{IAEA} were implemented in Geant4. The neutron escape probability depends on the target nucleus deexcitation model, which determines the neutron emission yield, as well as on the interaction cross-section of the emitted neutron with the crystal material. None of the available hadronic physics lists in Geant4 \cite{PhysicsList} gave good agreement with data. In particular, lists that used the G4NDL  library for high precision neutron physics clearly disagreed with the data. The optimal choice was found to be FTFP\_BERT\_PEN and photonuclear cross-sections scaled by a factor 1.5 from the encoded values in Geant4 to reproduce the intensity of the second photonuclear peak; the relevant peak for the tail fraction.
 
\vspace{0.1cm}

A correction to the expected amount of material traversed by positrons entering the calorimeter was required to match the position of the positron peak from $\pi \rightarrow e \nu$ decays in MC to that of the PIENU data as a function of positron angle.  It was concluded that there was less material at the entrance of the NaI(T$\ell$) crystal than given in the manufacturer’s specifications, which resulted in a $\sim$10\% correction for the amount of energy loss for positrons incident on the calorimeter.

At the three largest angles, the tail fraction given by the MC was significantly smaller than in the data. At these angles  shower leakages become important and the contribution from the energy deposited in the CsI to the total energy deposited is large. 
Plausible causes of the mismatch between MC and data are calibration errors of the CsI energy or missing material in the MC (irregular or unspecified shims and spacers). Reducing the CsI contribution by 8\% in the MC led to good agreement between MC and data at all angles.
\vspace{0.1cm}

%Number taken from 2021-2 p.95

To replicate the data digitization process and extraction of charge in the experimental setup, a routine was implemented in the MC. Waveform templates generated from the data were used and scaled to the simulated energy deposited in the calorimeter. The charge was obtained by integrating in the same time window as used in the data analysis. 

\begin{figure}[t!]
  \begin{center}
    \includegraphics[width=0.475\textwidth]{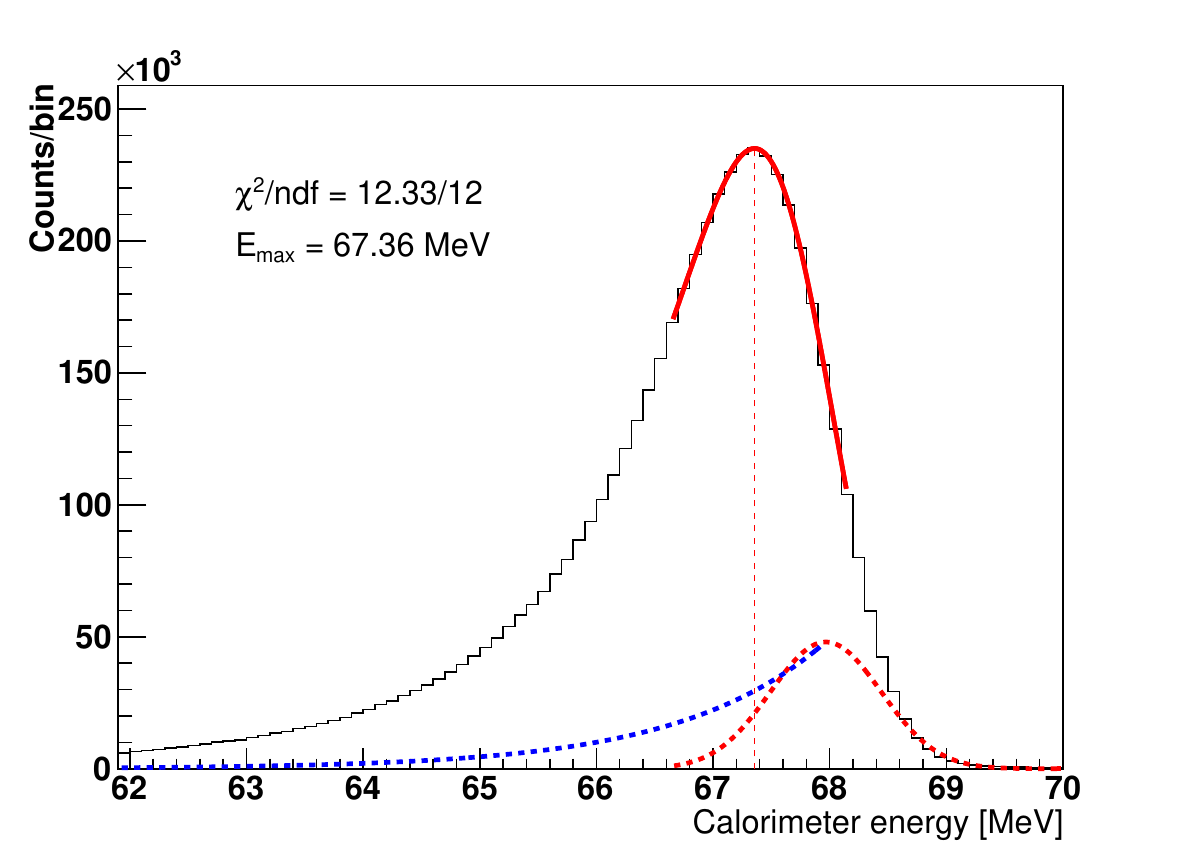}
    \caption{Fit for the positron peak of $\theta=0^\circ$ data. The fitting function consisted of a Gaussian convoluted with a falling exponential to mimic the energy loss due to the electromagnetic shower. Data are shown in black, the red solid line is the fit, the dashed blue line represents the falling exponential, and the dashed red line represents the Gaussian before convolution.}
  \label{fitted_peak}
  \end{center}
\end{figure}

\begin{figure}[t!]
  \begin{center}
    \subfloat{%
  \includegraphics[width=0.98\columnwidth]{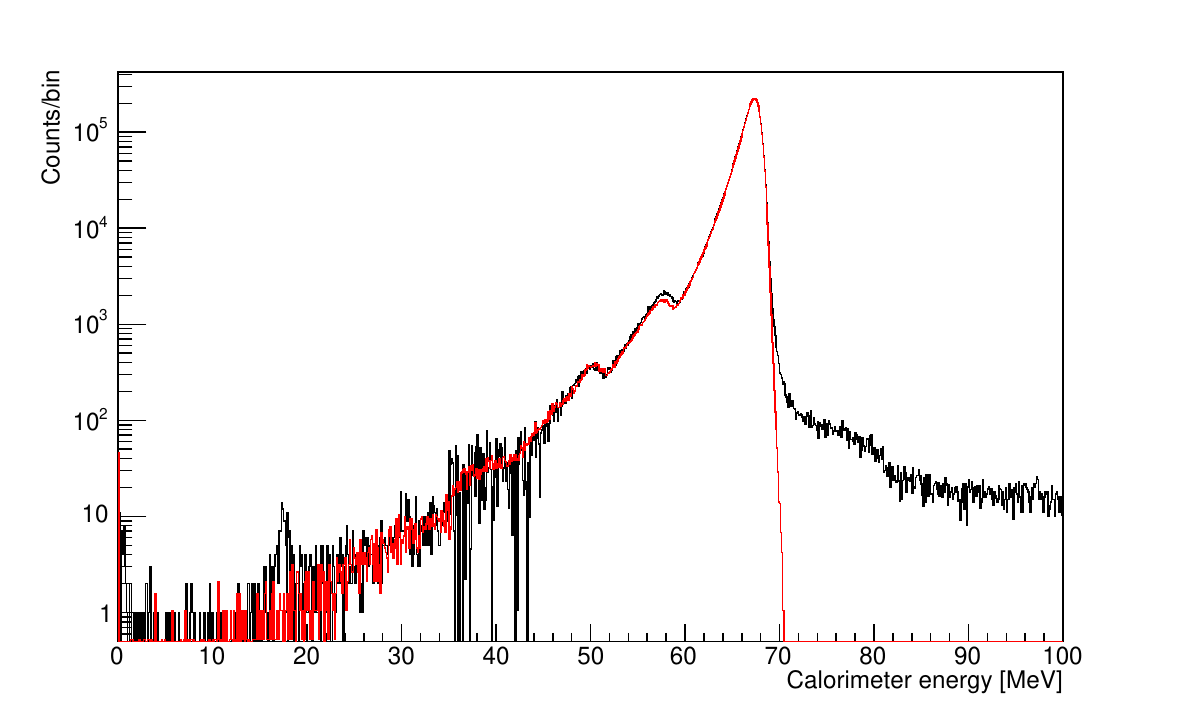}%
}\\
\subfloat{%
  \includegraphics[width=0.98\columnwidth]{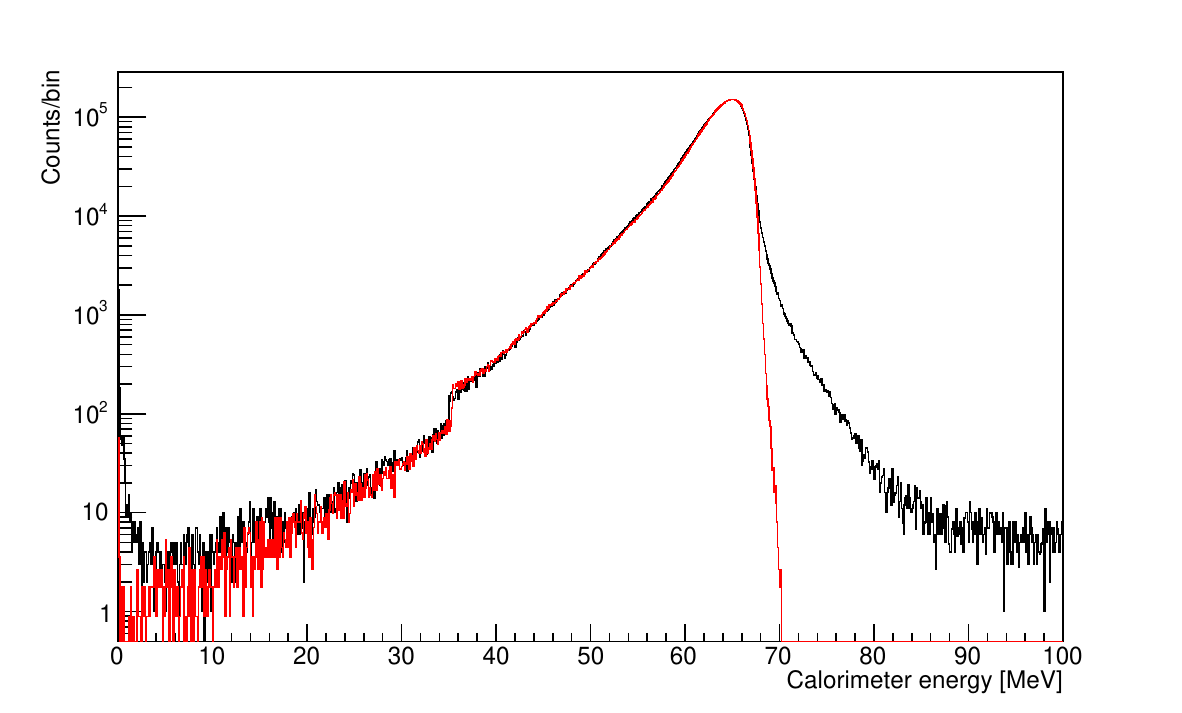}%
}
    \caption{Energy spectra of the positron beam in the calorimeter with  $\theta = 0^\circ$ (upper) and $\theta = 47.7^\circ$ (lower). Discontinuities at 35 MeV were due to the muon background subtraction described in the text. Data are shown in black and MC in red.}
  \label{positron_data_MC}
  \end{center}
\end{figure}

Since the data were uncalibrated, the data energy spectrum was normalized to the MC at each angle. To achieve this, the positron peaks for data and MC were fitted to a function as illustrated in Figure \ref{fitted_peak} for the $\theta=0^\circ$ data; the peak parameter was obtained with a precision of 10 keV. Figure \ref{positron_data_MC} shows the consistency between data and MC of the positron spectrum for $\theta=0^\circ$ and 47.7$^\circ$ in the upper and lower panels, respectively. The other eight angles had the same level of agreement between MC and data. The excellent agreement between data and MC is evidence against a low-momentum component of the positron beam; in fact, the data tail fraction for $\theta=0^\circ$ shows a deficit of $\sim$1\% relative to the MC.

To compare the tail fractions for MC to data, a possible mismatch of the data energy scale at the cut energy of 53.7 MeV contributed to the systematic uncertainty for the data.  The calorimeter energy scale for data was known to a precision of 0.1 MeV at the cut energy value.  The resultant systematic uncertainty is the dominant contribution to the values in Table \ref{systematic_muon} for data. Contributions to the systematic uncertainty for the MC tail fractions came from the adjustments to the MC enumerated above that were necessary to match the MC to the data. The largest of these came from the inability of the Geant4 models to reproduce the photonuclear peaks in the data. The results of these uncertainties (added in quadrature) are included in Table \ref{systematic_muon}.

\section{Conclusions}

\begin{figure}[h!]
  \begin{center}
    \includegraphics[width=0.48\textwidth]{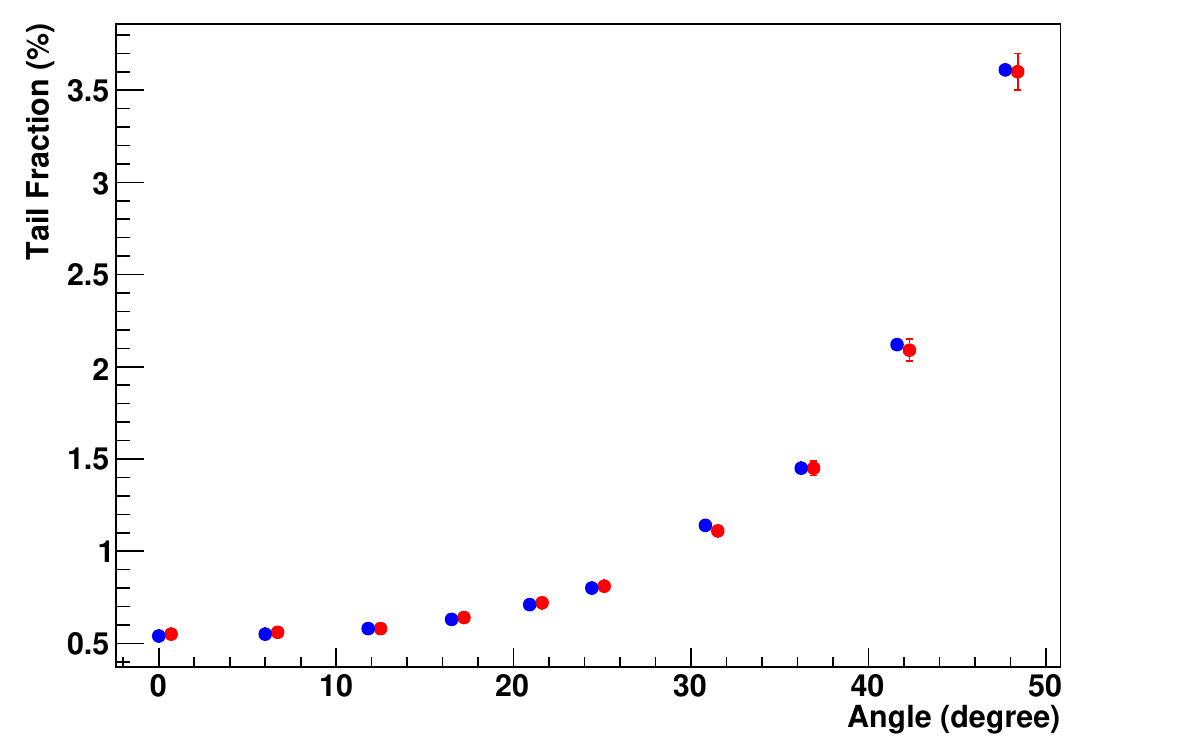}
    \caption{The PIENU  calorimeter tail fraction as a function of $\theta$ for 70 MeV/c positron data (blue) and MC (red). The statistical uncertainties were too small to be visible. In the plot, MC points were displaced by 0.7$^\circ$ on the abscissa to avoid overlapping with data.}
  \label{tail_mc_data}
  \end{center}
\end{figure}

The response function of the PIENU calorimeter was measured at ten angles using a positron beam. Figure \ref{tail_mc_data} shows the tail fraction as a function of the incidence angle of the beam with respect to the crystal axis in both data and MC. Those data were used to benchmark our MC simulation. In particular, EM physics lists and photo nuclear cross-sections were adjusted to reproduce the data most accurately at all angles. The level of agreement obtained supports the use of MC to reliably assess the tail fraction in the PIENU data.

%% bibitems, please use
%%

\section*{Acknowledgements}

This work was supported by the Natural Sciences and Engineering Research Council and TRIUMF through a contribution from the National Research Council of Canada, by CONAHCYT Mexico doctoral fellowship grants 312330 and 894777, and by JSPS KAKENHI Grants No. 21340059, and No. 19K03888 in Japan. We are grateful to Brookhaven National Laboratory for providing the NaI(T$\ell$) and CsI crystals. IH acknowledges support from the AANILD program of the Coordinación General de Estudios de Posgrado of UNAM, and from the PhD Exchange Program of the Arthur B. McDonald Institute.

\bibliographystyle{elsarticle-harv}

{}

\end{document}